\documentclass[12pt]{article}

\usepackage{epsfig}
\usepackage{amssymb}
\usepackage{amsbsy}
\usepackage{rotating}

\newcommand{\beq}{\begin{equation}}
\newcommand{\eeq}{\end{equation}}
\newcommand{\bea}{\begin{eqnarray}}
\newcommand{\eea}{\end{eqnarray}}
\newcommand{\nn}{\nonumber}

\def\eqn#1{Eq.~(\ref{#1})}

\def\fig#1{Fig.~{\ref{#1}}}

\def\cM{{\cal M}}
\def\sep{\mbox{$\,|\;$}}

\newcommand\sss{\scriptscriptstyle}
\newcommand\as{\alpha_{\sss S}}
\newcommand\mh{M_{\sss {\rm H}}}
\newcommand\mt{M_{t}}
\newcommand\pt{p_{{}\sss \perp}}
\newcommand\ph{p_{\sss {\rm H}}}
\newcommand\mhp{m_{\sss {{\rm H}_\perp}}}
\newcommand\sah{s_{j_1\sss {\rm H}}}
\newcommand\sbh{s_{j_2\sss {\rm H}}}
\newcommand\yh{y_{\sss {\rm H}}}
\newcommand\qip{q_{i\sss \perp}}

\begin{document}


\begin{center}
{\Large\bf The high-energy limit of $\boldsymbol{H+2}$ jet production via 

gluon fusion}
\vspace{1.cm}

{V.~Del~Duca$^1$, W.B.~Kilgore$^2$, C.~Oleari$^3$, C.R.~Schmidt$^4$
and D.~Zeppenfeld$^3$}\\
\vspace{.2cm}
{$^1$\sl I.N.F.N., Sezione di Torino\\
via P. Giuria, 1 - 10125 Torino, Italy}\\

\vspace{.2cm}
{$^2$\sl Department of Physics, 
Brookhaven National Laboratory\\
Upton, New York 11973-5000, USA}\\

\vspace{.2cm}
{$^3$\sl Department of Physics,
University of Wisconsin\\
Madison, WI 53706, USA}\\

\vspace{.2cm}
{$^4$\sl Department of Physics and Astronomy,
Michigan State University\\
East Lansing, MI 48824, USA}\\

\end{center}
\vspace{1cm}



At the Large Hadron Collider (LHC), the main production channels
of a Higgs boson are gluon fusion and weak-boson fusion 
(WBF)~\cite{CMS,ATLAS}. 
The WBF process, $q q\to q q H$, occurs through the exchange of a
$W$ or a $Z$ boson in the $t$ channel, and is characterized by
the production of two forward quark 
jets~\cite{Dokshitzer:1987nc}. 
Even though it is smaller than the gluon fusion 
channel by about a factor of 5 for an intermediate mass Higgs boson,
it is interesting because it is expected to provide information
on Higgs boson couplings~\cite{Zeppenfeld:2000td}. 
In this respect, $H + 2$ jet production
via gluon-gluon fusion, which has a larger production rate before cuts, can be 
considered a background; it
has the same final-state topology, and thus may hide the features of the
WBF process.

In Higgs production via gluon fusion, the Higgs boson is produced 
mostly via a top quark loop. The computation of $H + 2$ jet production
involves up to pentagon quark loops~\cite{DelDuca:2001eu}. However,
if the Higgs mass is smaller than the threshold for the creation of a 
top-quark pair, $\mh \lesssim 2 \mt$, the coupling of the Higgs to the
gluons via a top-quark loop can be replaced by an effective 
coupling~\cite{Shifman:1979eb}: this is called the {\it large-$\mt$ limit}.
It simplifies the calculation, because it reduces the number of loops 
in a given diagram by one. 
In $H + 2$ jet production, the large-$\mt$ limit yields a good
approximation to the exact calculation if, in addition to the
condition $\mh \lesssim 2 \mt$, we require that
the jet transverse energies are smaller than the top-quark mass, 
$\pt \lesssim \mt$~\cite{DelDuca:2001eu}.
However, 
the large $\mt$ approximation is quite insensitive to the value of the
Higgs--jet and/or dijet invariant masses. The last issue
is not academic, because Higgs
production via WBF, to which we should like to compare, features
typically two forward quark jets, and thus a large dijet invariant mass.

In this contribution, we consider $H + 2$ jet production 
when Higgs--jet and/or dijet 
invariant masses become much larger than the typical momentum 
transfers in the scattering. We term these conditions
the {\it high-energy limit}. 
In this limit the scattering amplitude factorizes into 
{\it impact factors} connected by a gluon exchanged in the $t$ channel.
Assembling together different impact factors, the amplitudes for different
sub-processes can be obtained. Thus the high-energy factorization 
constitutes a stringent consistency check on any amplitude for
the production of a Higgs plus one or more jets.

In the high-energy limit of $H + 2$ jet production,
the relevant (squared) energy scales are the parton center-of-mass energy $s$, 
the Higgs mass $\mh^2$, the dijet invariant mass $s_{j_1j_2}$, and 
the jet-Higgs invariant masses $\sah$ and $\sbh$. At leading order they
are related through momentum conservation,
\beq
s = s_{j_1j_2} + \sah + \sbh - \mh^2\, .\label{Hjjmtmcons}
\eeq
There are two possible high-energy limits to consider: 
$s_{j_1j_2}\gg\sah,\sbh\gg\mh^2$ and
$s_{j_1j_2},\sbh\gg\sah,\mh^2$.
In the first case the Higgs boson is centrally located in rapidity between
the two jets, and very far from either jet.  In the second case the Higgs boson
is close to one jet, say to jet $j_1$, in rapidity, and both of these are 
very far from jet $j_2$. In both cases
the amplitudes will factorize, and the relevant Higgs vertex in case 1
and the Higgs--gluon and Higgs--quark impact factors in case 2 can be obtained 
from the amplitudes for $q\,Q\to q\,Q\,H$ and $q\,g\to q\,g\,H$ scattering.

\subsection*{The high-energy limit $\boldsymbol{ s_{j_1j_2} \gg \sah,\, 
\sbh\gg \mh^2}$ }

We consider the production of two partons of momenta $p_1$ and $p_3$
and a Higgs boson of momentum $\ph$,
in the scattering between two partons of momenta $p_2$ and $p_4$, where 
all momenta are taken as outgoing.
We consider the limit in which the Higgs boson is produced 
centrally in rapidity, and very far from either jet,
$s_{j_1j_2} \gg \sah,\, \sbh\gg \mh^2$,
which is equivalent to require that
\beq
p_1^+ \gg \ph^+ \gg p_3^+ \, ,\qquad
p_1^- \ll \ph^- \ll p_3^- \, ,\label{mrk}
\eeq
where we have introduced the light-cone coordinates 
$p^{\pm}= p_0\pm p_z $, and complex transverse coordinates 
$p_{\perp} = p^x + i p^y$.
In the limit~(\ref{mrk}), the amplitudes
are dominated by gluon exchange in the $t$ channel, with emission
of the Higgs boson from the $t$-channel gluon. 
We can write the amplitude for $q\, Q\to q\, Q\, H$ scattering 
in the high-energy limit as~\cite{us}
\bea
\lefteqn{i\ \cM^{qq\to Hqq}(p_2^{-\nu_1},p_1^{\nu_1} \sep H \sep 
p_3^{\nu_3}, p_4^{-\nu_3}) } \nonumber\\ &=& 2s \left[ g\,
T^c_{a_1 \bar a_2}\, C^{\bar q;q}(p_2^{-\nu_1};p_1^{\nu_1})\right]
{1\over t_1} \left[ \delta^{cc'} C^{\sss H}(q_1,\ph,q_2)\right] {1\over t_2}
\left[ g\, T^{c'}_{a_3 \bar a_4}\, 
C^{\bar q;q}(p_4^{-\nu_3};p_3^{\nu_3})\right]\, ,\label{HqqqqHE}
\eea
where $q_1 = - (p_1+p_2)$, $q_2 = p_3+p_4$, $t_i\simeq - |\qip|^2$,
$i = 1, 2$, and the $\nu$'s are the quark helicities. 
In \eqn{HqqqqHE} we have made explicit the helicity conservation
along the massless quark lines.
The effective vertex $C^{\bar q;q}$ for the production of a quark jet,
$q\, g^* \rightarrow q$, 
contributes a phase factor~\cite{DelDuca:2000ha}: its square is 1.
The effective vertex for Higgs production
along the gluon ladder, $g^*\, g^* \rightarrow H$,
with $g^*$ an off-shell gluon, is
\begin{equation}
C^{\sss H}(q_1,\ph,q_2) = 2 {g^2 \mt^2\over v} \left( \mhp^2 
A_1(q_1,q_2) - 2A_2(q_1,q_2) \right)\, .\label{hif}
\end{equation} 
The scalar coefficients of the triangle vertex with two off-shell gluons,
$A_{1,2}$, are defined in terms of the form factors $F_T$ and $F_L$ of 
Ref.~\cite{DelDuca:2001eu} as
\beq
A_1 = {i\over (4\pi)^2} F_T\;, \qquad
A_2 = {i\over (4\pi)^2} \left( F_T\ q_1\cdot q_2 + F_L\ q_1^2 q_2^2
\right) \; .
\eeq

We have checked analytically that the amplitude for 
$q\, g\to q\, g\, H$ scattering can also be written as
\eqn{HqqqqHE}, provided we perform on one of the two effective vertices 
$C^{\bar q;q}$ the substitution (for the sake of illustration, we display 
it here for the lower vertex)
\begin{equation}
i g\, f^{bb'c}\, C^{g;g}(p_b^{\nu_b};p_{b'}^{\nu_{b'}}) \leftrightarrow g\, 
T^c_{b' \bar b}\, C^{\bar q;q}(p_b^{-\nu_{b'}};p_{b'}^{\nu_{b'}})\, 
,\label{qlrag}
\end{equation}
and use the effective vertices $g^*\, g \rightarrow g$ for the production
of a gluon jet~\cite{DelDuca:2000ha} (which contribute a phase factor as 
well). 
The same check on the (squared) amplitude for
$g\, g\to g\, g\, H$ scattering has been performed numerically.
Thus, in the high-energy limit~(\ref{mrk}), 
the amplitudes for $q\, Q\to q\, Q\, H$,
$q\, g\to q\, g\, H$ and $g\, g\to g\, g\, H$ scattering only
differ by the color strength in the jet-production vertex. Therefore, 
in a production rate it is enough to consider one of them and include 
the others through the effective parton distribution 
function~\cite{Combridge:1984jn},
$f_{\rm eff}(x,\mu_F^2) = G(x,\mu_F^2) + (C_F/C_A)\sum_f
\left[Q_f(x,\mu_F^2) + \bar Q_f(x,\mu_F^2)\right]$,
where $x$ is the momentum fraction of the incoming parton,
$\mu_F^2$ is the collinear factorization scale, 
and where the sum is over the quark flavors.

\begin{figure}[t]
\begin{center}
\begin{turn}{-90}
\epsfig{figure=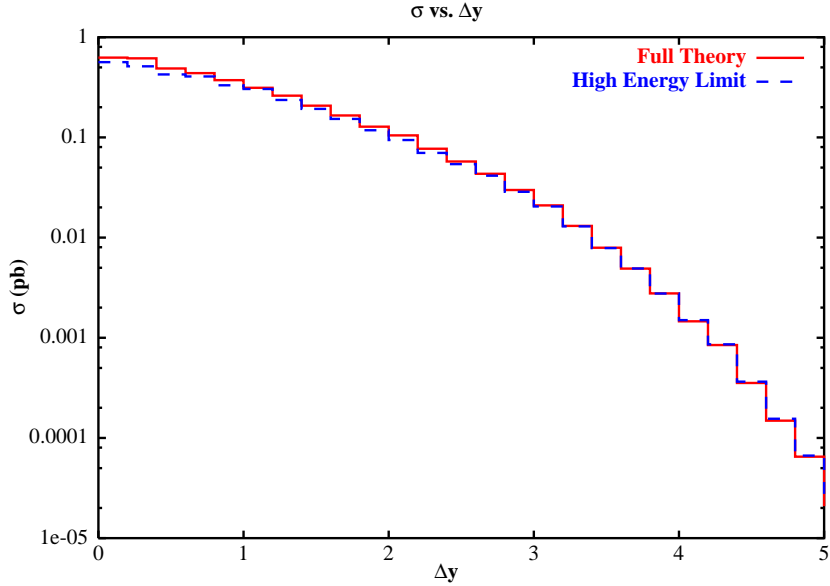,width=8cm}
\end{turn}
\end{center}
\caption{Cross section in $H+2$~jet production in $pp$ collisions 
at the LHC energy $\sqrt{s}=14$~TeV as a function of $\Delta y$,
with $\mh= 120$~GeV and $\mt = 175$~GeV. The dijet invariant mass
fulfills the constraint $\sqrt{s_{j_1j_2}} \ge 600$~GeV.
The rapidity interval $\Delta y$
is defined as $\Delta y = {\rm min}(|y_{j_1} - \yh|, |y_{j_2} - \yh|)$,
with the kinematical constraint $y_{j_1} > \yh > y_{j_2}$.
The solid line is the exact production rate; the dashed line is
the rate in the high-energy limit.} 
\label{fig:dydistra}
\end{figure}

In \fig{fig:dydistra} we plot the cross section in $H+2$~jet production 
in $pp$ collisions at the LHC energy $\sqrt{s}=14$~TeV, as a function 
of $\Delta y$, which is defined as the smallest rapidity difference 
between the Higgs and
the jets, $\Delta y = {\rm min}(|y_{j_1} - \yh|, |y_{j_2} - \yh|)$,
with the kinematical constraint $y_{j_1} > \yh > y_{j_2}$.
The solid line is the exact production rate,
with the amplitudes evaluated in Ref.~\cite{DelDuca:2001eu}; 
the dashed line is the rate in the high-energy limit~(\ref{mrk}),
with the amplitudes evaluated using Eqns.~(\ref{HqqqqHE})--(\ref{qlrag}).
It is apparent that the high-energy limit works very well over the whole
$\Delta y$ spectrum. However, in the evaluation of the effective 
vertex~(\ref{hif}), 
we used the exact value of the scalar coefficients $A_{1,2}$.
A more conservative statement is to say that when any kinematic quantity
involved in the amplitude~(\ref{HqqqqHE}) is evaluated in the 
limit~(\ref{mrk}),
we expect the high-energy limit to represent a good approximation of 
the exact calculation when $\Delta y\gtrsim 2$.

\subsection*{The high-energy limit 
$\boldsymbol{ s_{j_1j_2},\, \sbh\gg \sah,\, \mh^2}$ }
\label{sec:limit2}

Next, we consider the limit in which the Higgs boson is produced 
forward in rapidity, and close to one of the jets, say to jet $j_1$,
and both are very far from jet $j_2$, {\it i.e.}
$s_{j_1j_2},\, \sbh\gg \sah,\, \mh^2$. This limit implies that
\beq
p_1^+ \simeq \ph^+ \gg p_3^+ \, ,\qquad
p_1^- \simeq \ph^- \ll p_3^- \, .\label{mrk2}
\eeq
In this limit, the amplitudes are again dominated by gluon exchange in 
the $t$ channel, and factorize into in effective vertex for the
production of a jet and another for the production of a Higgs plus a jet.
For example, in the limit~(\ref{mrk2}) the amplitude for 
$q\ g\to q\ g\ H$ scattering~\cite{DelDuca:2001eu}
with the incoming gluon (quark) of momentum $p_2$ ($p_4$),
can be written as~\cite{us}
\bea
\lefteqn{ i\ \cM^{gq\to gHq}(p_2^{\nu_2}; p_1^{\nu_1}, H \sep 
p_3^{\nu_3}; p_4^{-\nu_3}) } \nn\\ &=& 2 s \left[i g\, f^{a_2a_1c}\, 
C^{g; {\sss H} g}(p_2^{\nu_2}; p_1^{\nu_1},\ph)\right] 
{1\over t} \left[i g\, T^c_{a_3 \bar a_4}\, 
C^{\bar q;q}(p_4^{-\nu_3};p_3^{\nu_3}) \right]\, ,\label{HqgqgHE}
\eea
where $C^{g; {\sss H} g}(p_2^{\nu_2}; p_1^{\nu_1},\ph)$ is the
effective vertex for the production of a Higgs boson and a gluon jet,
$g^* g\to g H$. It has two independent helicity configurations, which
we can take to be $C^{g; {\sss H} g}(p_2^-; p_1^+,\ph)$ and
$C^{g; {\sss H} g}(p_2^+; p_1^+,\ph)$~\cite{us}. 
High-energy factorization also implies that the amplitude for
$g\ g\to g\ g\ H$ scattering can be put in the form~(\ref{HqgqgHE}),
up to replacing the incoming quark with a gluon 
via the substitution~(\ref{qlrag}). Likewise,
the amplitude for $q\ Q\to q\ Q\ H$ scattering can be written as
\bea
\lefteqn{i\ \cM^{q Q\to q H Q}(p_2^{-\nu_1};p_1^{\nu_1}, \ph \sep 
p_3^{\nu_3}; p_4^{-\nu_3}) } \nonumber\\ &=& 
2s \left[ g\, T^c_{a_1 \bar a_2}\, 
C^{\bar q; {\sss H} q}(p_2^{-\nu_1};p_1^{\nu_1},\ph)\right] {1\over t}
\left[ g\, T^{c'}_{a_3 \bar a_4}\, 
C^{\bar q;q}(p_4^{-\nu_3};p_3^{\nu_3})\right]\, ,\label{HqqqqHE2}
\eea
where $C^{\bar q; {\sss H} q}(p_2^{-\nu_1};p_1^{\nu_1},\ph)$ is the 
effective vertex for the production of a Higgs and a quark jet,
$g^* q\to q H$. There is only one independent helicity configuration, which
we can take to be $C^{\bar q; {\sss H} q}(p_2^-;p_1^+,\ph)$, and its
expression
is given in Ref.~\cite{us}, where an analysis of the limit (\ref{mrk2})
with the kinematic parameters of \fig{fig:dydistra} can also be found.

In conclusion, we have considered $H + 2$ jet
production via gluon fusion, when either
one of the Higgs-jet or the dijet invariant masses become much larger
than the typical momentum transfers in the scattering. These limits
also occur naturally in Higgs production via WBF.
We have shown that
we can write the scattering amplitudes in accordance to high-energy
factorization, Eqns.~(\ref{HqqqqHE}), (\ref{HqgqgHE}) and (\ref{HqqqqHE2}).
The corresponding effective vertices, whose squares are the impact factors, 
can be found in Ref.~\cite{us}.


\begin{thebibliography}{99}

\bibitem{CMS}
G.~L.~Bayatian {\it et al.}, CMS Technical Proposal,
report CERN/LHCC/94-38x (1994);
R.~Kinnunen and D.~Denegri, 
{\it Expected SM/SUSY Higgs observability in CMS,}
CMS NOTE 1997/057;
R. Kinnunen and A. Nikitenko,
{\it Study of $H_{SUSY}\to \tau\tau\to l^{\pm}+h^{\mp} + E_t^{miss}$ in CMS,}
CMS TN/97-106;
R.~Kinnunen and D.~Denegri,
{\it The $H_{SUSY}\to \tau\tau\to h^{\pm}+h^{\mp}+X$
channel, its advantages and potential instrumental drawbacks,}
[hep-ph/9907291].

\bibitem{ATLAS}
ATLAS Collaboration, ATLAS TDR,
report CERN/LHCC/99-15 (1999).

\bibitem{Dokshitzer:1987nc}
Y.~L.~Dokshitzer, S.~I.~Troian and V.~A.~Khoze,
{\it Collective QCD Effects In The Structure Of Final Multi - Hadron States,}
Sov.\ J.\ Nucl.\ Phys.\  {\bf 46} (1987) 712.\\
J.~D.~Bjorken,
{\it Rapidity gaps and jets as a new physics signature in 
very high-energy hadron-hadron collisions,}
Phys.\ Rev.\  {\bf D47} (1993) 101.

\bibitem{Zeppenfeld:2000td}
D.~Zeppenfeld, R.~Kinnunen, A.~Nikitenko and E.~Richter-Was,
{\it Measuring Higgs boson couplings at the LHC,}
Phys.\ Rev.\ D {\bf 62} (2000) 013009 [hep-ph/0002036].

\bibitem{DelDuca:2001eu}
V.~Del Duca, W.~Kilgore, C.~Oleari, C.~Schmidt and D.~Zeppenfeld,
{\it H + 2 jets via gluon fusion,}
Phys.\ Rev.\ Lett.\  {\bf 87} (2001) 122001 [hep-ph/0105129].\\
V.~Del Duca, W.~Kilgore, C.~Oleari, C.~Schmidt and D.~Zeppenfeld,
{\it Gluon-fusion contributions to H + 2 jet production,}
Nucl.\ Phys.\ B {\bf 616} (2001) 367 [hep-ph/0108030].

\bibitem{Shifman:1979eb}
M.~A.~Shifman, A.~I.~Vainshtein, M.~B.~Voloshin and V.~I.~Zakharov,
{\it Low-Energy Theorems For Higgs Boson Couplings To Photons,}
Sov.\ J.\ Nucl.\ Phys.\  {\bf 30} (1979) 711.\\
J.~Ellis, M.~K.~Gaillard and D.~V.~Nanopoulos,
{\it A Phenomenological Profile Of The Higgs Boson,}
Nucl.\ Phys.\  {\bf B106} (1976) 292.

\bibitem{us}
V.~Del Duca, W.~Kilgore, C.~Oleari, C.~Schmidt and D.~Zeppenfeld,
in preparation.

\bibitem{DelDuca:2000ha}
V.~Del Duca, A.~Frizzo and F.~Maltoni,
{\it Factorization of tree QCD amplitudes in the high-energy limit 
and in  the collinear limit,}
Nucl.\ Phys.\  {\bf B568} (2000) 211 [hep-ph/9909464].

\bibitem{Combridge:1984jn}
B.~L.~Combridge and C.~J.~Maxwell,
{\it Untangling Large P(T) Hadronic Reactions,}
Nucl.\ Phys.\  {\bf B239} (1984) 429.

\end{thebibliography}
\end{document}